\documentclass[10pt,a4paper,onecolumn]{article}
\usepackage{marginnote}
\usepackage{graphicx}
\usepackage{xcolor}
\usepackage{authblk,etoolbox}
\usepackage{titlesec}
\usepackage{calc}
\usepackage{tikz}
\usepackage{hyperref}
\hypersetup{colorlinks,breaklinks=true,
            urlcolor=[rgb]{0.0, 0.5, 1.0},
            linkcolor=[rgb]{0.0, 0.5, 1.0}}
\usepackage{caption}
\usepackage{tcolorbox}
\usepackage{amssymb,amsmath}
\usepackage{ifxetex,ifluatex}
\usepackage{seqsplit}
\usepackage{xstring}

\usepackage{float}
\let\origfigure\figure
\let\endorigfigure\endfigure
\renewenvironment{figure}[1][2] {
    \expandafter\origfigure\expandafter[H]
} {
    \endorigfigure
}

\usepackage{fixltx2e} 
\usepackage[
  backend=biber,
]{biblatex}
\bibliography{paper.bib}

\let\textttOrig=\texttt
\def\texttt#1{\expandafter\textttOrig{\seqsplit{#1}}}
\renewcommand{\seqinsert}{\ifmmode
  \allowbreak
  \else\penalty6000\hspace{0pt plus 0.02em}\fi}

\makeatletter
\let\href@Orig=\href
\def\href@Urllike#1#2{\href@Orig{#1}{\begingroup
    \def\Url@String{#2}\Url@FormatString
    \endgroup}}
\def\href@Notdoi#1#2{\def\tempa{#1}\def\tempb{#2}%
  \ifx\tempa\tempb\relax\href@Urllike{#1}{#2}\else
  \href@Orig{#1}{#2}\fi}
\def\href#1#2{%
  \IfBeginWith{#1}{https://doi.org}%
  {\href@Urllike{#1}{#2}}{\href@Notdoi{#1}{#2}}}
\makeatother

\newlength{\cslhangindent}
\newlength{\csllabelwidth}
\newenvironment{CSLReferences}[3] 
 {
  \setlength{\parindent}{0pt}
  \ifodd #1 \everypar{\setlength{\hangindent}{\cslhangindent}}\ignorespaces\fi
  \ifnum #2 > 0
  \setlength{\parskip}{#2\baselineskip}
  \fi
 }%
 {}
\usepackage{calc}

\usepackage[top=3.5cm, bottom=3cm, right=1.5cm, left=1.0cm,
            headheight=2.2cm, reversemp, includemp, marginparwidth=4.5cm]{geometry}

\titleformat{\section}
  {\normalfont\sffamily\Large\bfseries}
  {}{0pt}{}
\titleformat{\subsection}
  {\normalfont\sffamily\large\bfseries}
  {}{0pt}{}
\titleformat{\subsubsection}
  {\normalfont\sffamily\bfseries}
  {}{0pt}{}
\titleformat*{\paragraph}
  {\sffamily\normalsize}

\usepackage{fancyhdr}
\makeatletter
\let\ps@plain\ps@fancy
\fancyheadoffset[L]{4.5cm}
\fancyfootoffset[L]{4.5cm}

\definecolor{linky}{rgb}{0.0, 0.5, 1.0}

\newtcolorbox{repobox}
   {colback=red, colframe=red!75!black,
     boxrule=0.5pt, arc=2pt, left=6pt, right=6pt, top=3pt, bottom=3pt}

\newcommand{\ExternalLink}{%
   \tikz[x=1.2ex, y=1.2ex, baseline=-0.05ex]{%
       \begin{scope}[x=1ex, y=1ex]
           \clip (-0.1,-0.1)
               --++ (-0, 1.2)
               --++ (0.6, 0)
               --++ (0, -0.6)
               --++ (0.6, 0)
               --++ (0, -1);
           \path[draw,
               line width = 0.5,
               rounded corners=0.5]
               (0,0) rectangle (1,1);
       \end{scope}
       \path[draw, line width = 0.5] (0.5, 0.5)
           -- (1, 1);
       \path[draw, line width = 0.5] (0.6, 1)
           -- (1, 1) -- (1, 0.6);
       }
   }

\patchcmd{\@maketitle}{center}{flushleft}{}{}
\patchcmd{\@maketitle}{center}{flushleft}{}{}
\patchcmd{\@maketitle}{\LARGE}{\LARGE\sffamily}{}{}
\def\maketitle{{%
  
  \AB@maketitle}}
\makeatletter
\renewcommand\AB@affilsepx{ \protect\Affilfont}
\renewcommand\AB@affilnote[1]{{\bfseries #1}\hspace{3pt}}
\renewcommand{\affil}[2][]%
   {\newaffiltrue\let\AB@blk@and\AB@pand
      \if\relax#1\relax\def\AB@note{\AB@thenote}\else\def\AB@note{#1}%
        \setcounter{Maxaffil}{0}\fi
        \begingroup
        \let\href=\href@Orig
        \let\texttt=\textttOrig
        \let\protect\@unexpandable@protect
        \def\thanks{\protect\thanks}\def\footnote{\protect\footnote}%
        \@temptokena=\expandafter{\AB@authors}%
        {\def\\{\protect\\\protect\Affilfont}\xdef\AB@temp{#2}}%
         \xdef\AB@authors{\the\@temptokena\AB@las\AB@au@str
         \protect\\[\affilsep]\protect\Affilfont\AB@temp}%
         \gdef\AB@las{}\gdef\AB@au@str{}%
        {\def\\{, \ignorespaces}\xdef\AB@temp{#2}}%
        \@temptokena=\expandafter{\AB@affillist}%
        \xdef\AB@affillist{\the\@temptokena \AB@affilsep
          \AB@affilnote{\AB@note}\protect\Affilfont\AB@temp}%
      \endgroup
       \let\AB@affilsep\AB@affilsepx
}
\makeatother

\renewcommand\Affilfont{\sffamily\small\mdseries}
\ifnum 0\ifxetex 1\fi\ifluatex 1\fi=0 
  \usepackage[T1]{fontenc}
  \usepackage[utf8]{inputenc}

\else 
  \ifxetex
    \usepackage{mathspec}
    \usepackage{fontspec}

  \else
    \usepackage{fontspec}
  \fi
  \defaultfontfeatures{Ligatures=TeX,Scale=MatchLowercase}

\fi
\IfFileExists{upquote.sty}{\usepackage{upquote}}{}
\IfFileExists{microtype.sty}{%
\usepackage{microtype}
\UseMicrotypeSet[protrusion]{basicmath} 
}{}

\usepackage{hyperref}
\hypersetup{unicode=true,
            pdftitle={ExoTiC-LD: thirty seconds to stellar limb-darkening coefficients},
            pdfborder={0 0 0},
            breaklinks=true}
\let\addcontentslineOrig=\addcontentsline
\def\addcontentsline#1#2#3{\bgroup
  \let\texttt=\textttOrig\addcontentslineOrig{#1}{#2}{#3}\egroup}
\let\markbothOrig\markboth
\def\markboth#1#2{\bgroup
  \let\texttt=\textttOrig\markbothOrig{#1}{#2}\egroup}
\let\markrightOrig\markright
\def\markright#1{\bgroup
  \let\texttt=\textttOrig\markrightOrig{#1}\egroup}

\IfFileExists{parskip.sty}{%
\usepackage{parskip}
}{
\setlength{\parindent}{0pt}
\setlength{\parskip}{6pt plus 2pt minus 1pt}
}
\ifx\paragraph\undefined\else
\let\oldparagraph\paragraph
\renewcommand{\paragraph}[1]{\oldparagraph{#1}\mbox{}}
\fi
\ifx\subparagraph\undefined\else
\let\oldsubparagraph\subparagraph
\renewcommand{\subparagraph}[1]{\oldsubparagraph{#1}\mbox{}}
\fi

\title{ExoTiC-LD: thirty seconds to stellar limb-darkening coefficients}

        \author[1]{David Grant}
        \author[1]{Hannah R. Wakeford}
    
      \affil[1]{HH Wills Physics Laboratory, University of Bristol, Tyndall Avenue, Bristol, BS8 1TL, UK}
      
  \date{\vspace{-0ex}}

\begin{document}
\maketitle

\marginpar{

  \begin{flushleft}
  \sffamily\small

  {\bfseries DOI:} \href{https://doi.org/10.21105/joss.06816}{\color{linky}{10.21105/joss.06816}}

  \vspace{2mm}

  {\bfseries Software}
  \begin{itemize}
    \setlength\itemsep{0em}
    \item \href{https://github.com/openjournals/joss-reviews/issues/6816}{\color{linky}{Review}} \ExternalLink
    \item \href{https://github.com/Exo-TiC/ExoTiC-LD}{\color{linky}{Repository}} \ExternalLink
    \item \href{https://doi.org/10.5281/zenodo.13224465}{\color{linky}{Archive}} \ExternalLink
  \end{itemize}

  \vspace{2mm}

  \par\noindent\hrulefill\par

  \vspace{2mm}

  {\bfseries Editor:} \href{https://warrickball.gitlab.io/}{Warrick Ball} \ExternalLink \\
  \vspace{1mm}
    {\bfseries Reviewers:}
  \begin{itemize}
  \setlength\itemsep{0em}
    \item \href{https://github.com/nenasedk}{@nenasedk}
    \item \href{https://github.com/LorenzoMugnai}{@LorenzoMugnai}
    \end{itemize}
    \vspace{2mm}

  {\bfseries Submitted:} 23 April 2024\\
  {\bfseries Published:} 09 August 2024

  \vspace{2mm}
  {\bfseries License}\\
  Authors of papers retain copyright and release the work under a Creative Commons Attribution 4.0 International License (\href{http://creativecommons.org/licenses/by/4.0/}{\color{linky}{CC BY 4.0}}).

  \end{flushleft}
}

\hypertarget{summary}{%
\section{Summary}\label{summary}}

Stellar limb darkening is the observed variation in brightness of a star between its centre and edge (or limb) when viewed in the plane of the sky. Stellar brightness is maximal at the centre and then decreases radially and monotonically towards the limb – hence the term “limb darkening”. This effect is crucial for finding and characterising planets beyond our Solar System, known as exoplanets, as these planets are often studied when crossing in front of their host stars. As such, limb darkening is directly linked to the exoplanet signals. Limb darkening is typically modelled by one of various functional forms, as outlined in Claret (\href{ref-Albert2023}{2000}) and Sing (\href{ref-Sing}{2010}), and the coefficients of these functions is what ExoTiC-LD is designed to compute. A wide variety of functional forms are supported, including those benchmarked by Espinoza \& Jordán (\href{ref-Espinoza}{2016}) as well as reparameterisations suggested by Kipping (\href{ref-Kipping}{2013}).

\hypertarget{statement-of-need}{%
\section{Statement of Need}\label{statement-of-need}}

Stellar limb darkening depends on the type of star, the wavelengths of light being observed, and the sensitivity of the instrument/telescope performing the observation. Therefore, to compute limb-darkening coefficients requires a frustrating amount of “data admin”. In brief, one starts with a search through grids of stellar models to find a good match with the science target in metallicity, effective temperature, and surface gravity. Then, one must retrieve the wavelength-dependent sensitivity of the employed instrument, process all these data into a cohesive form, and then finally compute the limb-darkening coefficients.

Previous software has made calculating limb-darkening coefficients available to the community (e.g., Bourque et al., \href{ref-Bourque}{2021}; Morello et al., \href{ref-Morello}{2020}; Parviainen \& Aigrain, \href{ref-Parviainen}{2015}; Southworth, \href{ref-Southworth}{2008}), albeit with varying degrees of installation complexity and access to stellar and instrument data. In ExoTiC-LD we have done all of the heavy lifting for the user, making the process as fast and frictionless as possible. A user simply has to pip install the code and the relevant data will be automatically downloaded at runtime and the limb-darkening coefficients computed. In particular, a wide selection of stellar and instrument data has been pre-processed and homogenised. Additionally, the stellar model grids have been stored as tree structures, enabling an efficient search for good matches and helpful warnings to the user. Currently, the stellar models supported are PHOENIX (Husser et al., \href{ref-Husser}{2013}), kurucz (Kurucz, \href{ref-Bourque}{1993}), stagger (Magic et al., \href{ref-Magic}{2015}), and MPS-ATLAS (NM Kostogryz et al., \href{ref-Kostogryz-NM}{2022}; N. Kostogryz et al., \href{ref-Kostogryz-N}{2023}). There are also options to provide custom data if the user has their own stellar models or instrument
data.

ExoTiC-LD thus far has predominantly been utilised in the study of exoplanet atmospheres, helping to facilitate the study of Jupiter-like (e.g., Alderson et al., \href{ref-Alderson}{2023}; Grant et al., \href{ref-Grant}{2023}), Neptune-like (e.g., Radica et al., \href{ref-Radica}{2024}; Roy et al., \href{ref-Roy}{2023}), and Earth-like exoplanets (e.g., Kirk et al., \href{ref-Kirk}{2024}; Moran et al., \href{ref-Moran}{2023}). It has also been incorporated into the popular open-source JWST data reduction and analysis pipeline, called Eureka! (Bell et al., \href{ref-Bell}{2022}).

\hypertarget{acknowledgements}{%
\section{Acknowledgements}\label{acknowledgements}}

We acknowledge contributions from Natasha Batalha, Matthew Hill, and Iva Laginja, as well as testing by Taylor Bell, Lili Alderson, Daniel Valentine, Charlotte Fairman, Katy Chubb, and Nikole Lewis. D.G. and H.R.W were funded by UK Research and Innovation (UKRI) under the UK government’s Horizon Europe funding guarantee as part of an ERC Starter Grant [grant number EP/Y006313/1].

\hypertarget{references}{%
\section*{References}\label{references}}
\addcontentsline{toc}{section}{References}

\hypertarget{refs}{}
\begin{CSLReferences}{1}{0}

\leavevmode\hypertarget{ref-Alderson}{}%
Alderson, L., Wakeford, H. R., Alam, M. K., Batalha, N. E., Lothringer, J. D., Adams Redai, J., Barat, S., Brande, J., Damiano, M., Daylan, T., Espinoza, N., Flagg, L., Goyal, J. M., Grant, D., Hu, R., Inglis, J., Lee, E. K. H., Mikal-Evans, T., Ramos-Rosado, L., … Zhang, X. (2023). {Early Release Science of the exoplanet WASP-39b with JWST NIRSpec G395H.} \emph{Nature}, \emph{614}(7949), 664–669. \url{https://doi.org/10.1038/s41586-022-05591-3}

\leavevmode\hypertarget{ref-Bell}{}%
Bell, T., Ahrer, E.-M., Brande, J., Carter, A., Feinstein, A., Caloca, G., Mansfield, M., Zieba, S., Piaulet, C., Benneke, B., Filippazzo, J., May, E., Roy, P.-A., Kreidberg, L., \& Stevenson, K. (2022). {Eureka!: An End-to-End Pipeline for JWST Time-Series Observations.} \emph{The Journal of Open Source Software}, \emph{7}(79), 4503. \url{https://doi.org/10.21105/joss.04503}

\leavevmode\hypertarget{ref-Bourque}{}%
Bourque, M., Espinoza, N., Filippazzo, J., Fix, M., King, T., Martlin, C., Medina, J., Batalha, N., Fox, M., Fowler, J., Fraine, J., Hill, M., Lewis, N., Stevenson, K., Valenti, J., \& Wakeford, H. (2021). {The exoplanet characterization toolkit (ExoCTK)} (Version 1.0.0).
\emph{Zenodo}. \url{https://doi.org/10.5281/zenodo.4556063}

\leavevmode\hypertarget{ref-Claret}{}%
Claret, A. (2000). {A new non-linear limb-darkening law for LTE stellar atmosphere models. Calculations for -5.0 <= log[M/H] <= +1, 2000 K <= $T_{\rm{eff}}$ <= 50000 K at several surface gravities.} \emph{Astronomy \& Astrophysics}, \emph{363}, 1081–1190.

\leavevmode\hypertarget{ref-Espinoza}{}%
Espinoza, N., \& Jordán, A. (2016). {Limb darkening and exoplanets - II. Choosing the best law for optimal retrieval of transit parameters.} \emph{MNRAS}, \emph{457}(4), 3573–3581. \url{https://doi.org/10.1093/mnras/stw224}

\leavevmode\hypertarget{ref-Grant}{}%
Grant, D., Lewis, N. K., Wakeford, H. R., Batalha, N. E., Glidden, A., Goyal, J., Mullens, E., MacDonald, R. J., May, E. M., Seager, S., Stevenson, K. B., Valenti, J. A., Visscher, C., Alderson, L., Allen, N. H., Cañas, C. I., Colón, K., Clampin, M., Espinoza, N., … Watkins, L. L. (2023). {JWST-TST DREAMS: Quartz Clouds in the Atmosphere of WASP-17b.} \emph{Astrophysical Journal Letters}, \emph{956}(2), L29. \url{https://doi.org/10.3847/2041-8213/acfc3b}

\leavevmode\hypertarget{ref-Husser}{}%
Husser, T.-O., Wende-von Berg, S., Dreizler, S., Homeier, D., Reiners, A., Barman, T., \& Hauschildt, P. H. (2013). {A new extensive library of PHOENIX stellar atmospheres and synthetic spectra.} \emph{Astronomy \& Astrophysics}, \emph{553}. \url{https://doi.org/10.1051/0004-6361/201219058}

\leavevmode\hypertarget{ref-Kipping}{}%
Kipping, D. M. (2013). {Efficient, uninformative sampling of limb darkening coefficients for two-parameter laws.} \emph{MNRAS}, \emph{435}(3), 2152–2160. \url{https://doi.org/10.1093/mnras/stt1435}

\leavevmode\hypertarget{ref-Kirk}{}%
Kirk, J., Stevenson, K. B., Fu, G., Lustig-Yaeger, J., Moran, S. E., Peacock, S., Alam, M. K., Batalha, N. E., Bennett, K. A., Gonzalez-Quiles, J., López-Morales, M., Lothringer, J. D., MacDonald, R. J., May, E. M., Mayorga, L. C., Rustamkulov, Z., Sing, D. K., Sotzen, K. S., Valenti, J. A., \& Wakeford, H. R. (2024). {JWST/NIRCam Transmission Spectroscopy of the Nearby Sub-Earth GJ 341b.} \emph{Astrophysical Journal}, \emph{167}(3), 90. \url{https://doi.org/10.3847/1538-3881/ad19df}

\leavevmode\hypertarget{ref-Kostogryz-N}{}%
Kostogryz, N., Shapiro, A., Witzke, V., Grant, D., Wakeford, H., Stevenson, K., Solanki, S., \& Gizon, L. (2023). {MPS-ATLAS library of stellar model atmospheres and spectra.} \emph{Research Notes of the AAS}, \emph{7}(3), 39. \url{https://doi.org/10.3847/2515-5172/acc180}

\leavevmode\hypertarget{ref-Kostogryz-NM}{}%
Kostogryz, NM, Witzke, V., Shapiro, A., Solanki, S., Maxted, P., Kurucz, R., \& Gizon, L. (2022). {Stellar limb darkening. A new MPS-ATLAS library for kepler, TESS, CHEOPS, and PLATO passbands.} \emph{Astronomy \& Astrophysics}, \emph{666}, A60. \url{https://doi.org/10.1051/0004-6361/202243722}

\leavevmode\hypertarget{ref-Kurucz}{}%
Kurucz, R.-L. (1993). {ATLAS9 stellar atmosphere programs and 2km/s grid.} \emph{Kurucz CD-Rom}, \emph{13}.

\leavevmode\hypertarget{ref-Magic}{}%
Magic, Z., Chiavassa, A., Collet, R., \& Asplund, M. (2015). {The stagger-grid: A grid of 3D stellar atmosphere models-IV. Limb darkening coefficients.} \emph{Astronomy \& Astrophysics}, \emph{573}, A90. \url{https://doi.org/10.1051/0004-6361/201423804}

\leavevmode\hypertarget{ref-Moran}{}%
Moran, S. E., Stevenson, K. B., Sing, D. K., MacDonald, R. J., Kirk, J., Lustig-Yaeger, J., Peacock, S., Mayorga, L. C., Bennett, K. A., López-Morales, M., May, E. M., Rustamkulov, Z., Valenti, J. A., Adams Redai, J. I., Alam, M. K., Batalha, N. E., Fu, G., GonzalezQuiles, J., Highland, A. N., … Wakeford, H. R. (2023). {High Tide or Riptide on the Cosmic Shoreline? A Water-rich Atmosphere or Stellar Contamination for the Warm Super-Earth GJ 486b from JWST Observations.} \emph{Astrophysical Journal Letters}, \emph{948}(1), L11. \url{https://doi.org/10.3847/2041-8213/accb9c}

\leavevmode\hypertarget{ref-Morello}{}%
Morello, G., Claret, A., Martin-Lagarde, M., Cossou, C., Tsiara, A., \& Lagage, P.-O. (2020). {ExoTETHyS: Tools for exoplanetary transits around host stars.} \emph{Journal of Open Source Software}, \emph{5}(46), 1834. \url{https://doi.org/10.21105/joss.01834}

\leavevmode\hypertarget{ref-Parviainen}{}%
Parviainen, H., \& Aigrain, S. (2015). {ldtk: Limb Darkening Toolkit.} \emph{MNRAS}, \emph{453}(4), 3821–3826. \url{https://doi.org/10.1093/mnras/stv1857}

\leavevmode\hypertarget{ref-Radica}{}%
Radica, M., Coulombe, L.-P., Taylor, J., Albert, L., Allart, R., Benneke, B., Cowan, N. B., Dang, L., Lafrenière, D., Thorngren, D., Artigau, É., Doyon, R., Flagg, L., Johnstone, D., Pelletier, S., \& Roy, P.-A. (2024). {Muted Features in the JWST NIRISS Transmission Spectrum of Hot Neptune LTT 9779b.} \emph{Astrophysical Journal Letters}, \emph{962}(1), L20. \url{https://doi.org/10.3847/2041-8213/ad20e4}

\leavevmode\hypertarget{ref-Roy}{}%
Roy, P.-A., Benneke, B., Piaulet, C., Gully-Santiago, M. A., Crossfield, I. J. M., Morley, C. V., Kreidberg, L., Mikal-Evans, T., Brande, J., Delisle, S., Greene, T. P., Hardegree-Ullman, K. K., Barman, T., Christiansen, J. L., Dragomir, D., Fortney, J. J., Howard, A. W., Kosiarek, M. R., \& Lothringer, J. D. (2023). {Water Absorption in the Transmission Spectrum of the Water World Candidate GJ 9827 d.} \emph{Astrophysical Journal Letters}, \emph{954}(2), L52. \url{https://doi.org/10.3847/2041-8213/acebf0}

\leavevmode\hypertarget{ref-Sing}{}%
Sing, D. K. (2010). {Stellar limb-darkening coefficients for CoRot and Kepler.} \emph{Astronomy \& Astrophysics}, \emph{510}, A21. \url{https://doi.org/10.1051/0004-6361/200913675}

\leavevmode\hypertarget{ref-Southworth}{}%
Southworth, J. (2008). {Homogeneous studies of transiting extrasolar planets - I. Light-curve analyses.} \emph{MNRAS}, \emph{386}(3), 1644–1666. \url{https://doi.org/10.1111/j.1365-2966.2008.13145.x}

\end{CSLReferences}

\end{document}